\documentclass{aa}  

\usepackage{graphicx}
\usepackage{txfonts}

\usepackage{color}

\begin{document}

\title{Detailed study of SNR G306.3--0.9 using XMM-{\em Newton} and {\em Chandra} observations}

\author{J.A. Combi\inst{1,2}, F. Garc\'{\i}a\inst{1,2}, A.E. Su\'arez\inst{1,2}, P.L. Luque-Escamilla\inst{3}, S. Paron\inst{4,5}, M. Miceli\inst{6,7}}

\authorrunning{Combi et al.}

\titlerunning{Detailed X-ray study of SNR G306.3--0.9} 

\offprints{J. Combi}

\institute{Instituto Argentino de Radioastronom\'{\i}a (CCT La Plata, CONICET), C.C.5, (1894) Villa Elisa, Buenos Aires, Argentina.\\
 \email{[jcombi,fgarcia,suarez]@iar-conicet.gov.ar}
 \and
Facultad de Ciencias Astron\'omicas y Geof\'{\i}sicas, Universidad Nacional de La Plata, Paseo del Bosque, B1900FWA La Plata, Argentina.
\and
Deptamento de Ingenier\'ia Mec\'anica y Minera, Universidad de Ja\'en, Campus Las Lagunillas s/n Ed. A3 Ja\'en, Spain, 23071.\\
\email{peter@ujaen.es}
\and
Instituto de Astronom\'ia y F\'isica del Espacio (IAFE), CC 67, Suc. 28, 1428 Buenos Aires, Argentina.  
\and 
FADU and CBC, Universidad de Buenos Aires, Buenos Aires, Argentina.
\and
Dipartimento di Fisica \& Chimica, Universit\`a di Palermo, Piazza del Parlamento 1, I-90134 Palermo, Italy.
\and
INAF-Osservatorio Astronomico di Palermo, P.zza del Parlamento 1, I-90134 Palermo, Italy.}

   \date{Received; accepted}

 
  \abstract
   {}
   {We aim to study the spatial distribution of the physical and chemical properties of the X-ray emitting plasma of the supernova remnant (SNR) G306.3$-$0.9 in detail to obtain constraints on its ionization stage, the progenitor supernova explosion, and the age of the remnant.}
   {We used combined data from {\it XMM-Newton} and {\it Chandra} observatories to study the X-ray morphology of G306.3$-$0.9 in detail.  A spatially resolved spectral analysis was used to obtain physical and geometrical parameters of different regions of the remnant. {\it Spitzer} infrared observations, available in the archive, were also used to constrain the progenitor supernova and study the environment in which the remnant evolved.}
        {The X-ray morphology of the remnant displays a non-uniform structure of semi-circular appearance, with a bright southwest region and very weak or almost negligible X-ray emission in its northern part. These results indicate that the remnant is propagating in a non-uniform environment as the shock fronts are encountering a high-density medium, where enhanced infrared emission is detected. The X-ray spectral analysis of the selected regions shows distinct emission-line features of several metal elements, confirming the thermal origin of the emission. The X-ray spectra are well represented by a combination of two absorbed thermal plasma models: one in equilibrium ionization (VAPEC) with a mean temperature of $\sim$0.19 keV, and another out of equilibrium ionization (VNEI) at a higher temperature of $\sim$1.1 or 1.6--1.9 keV. For regions located in the northeast, central, and southwest part of the SNR, we found elevated abundances of Si, S, Ar, Ca, and Fe, typical of ejecta material. The outer regions located northwest and south show values of the abundances above solar but lower than to those found in the central regions. This suggests that the composition of the emitting outer parts of the SNR is a combination of ejecta and shocked material of the interstellar medium. The comparison between the S/Si, Ar/Si, and Ca/Si abundances ratios (1.75, 1.27, and 2.72 in the central region, respectively), favor a Type Ia progenitor for this remnant, a result that is also supported by an independent morphological analysis using the X-ray and 24~$\mu$m IR data.}



   \maketitle
%

\section{Introduction}

The energetic phenomena of a supernova (SN) explosion produce a very profound impact on the interstellar medium (ISM) because it imparts a  large amount of mechanical energy and heavy elements to the surrounding medium. The interaction of the shock-front of supernova remnants (SNRs) with the environment greatly affects the dynamic structure of the Galaxy. As a result, we observe SNRs with different types of morphologies at radio, infrared  (IR), and X-ray wavelengths. The study of the SNR morphology at infrared and X-ray frequencies can be used as a powerful observational diagnostic to distinguish between progenitor types and to improve our knowledge of the physical and chemical properties of the X-ray emitting plasma.

\begin{table*}
\caption{{\it Chandra} and XMM-{\it Newton} observations of G306.3-0.9.}
\label{obs}\centering
\begin{center}
\begin{tabular}{l c c l c c c c}
\hline\hline      
Satellite& \multicolumn{2}{c}{{\it Chandra}}&& \multicolumn{1}{c}{XMM-{\it Newton}} \\ 
\cline{2-3} \cline{5-6}
Obs-Id                   & 13419 & 14812 && 0691550101 \\ 
Date                    & 13/02/2012  & 13/12/2014      && 03/02/2013 \\
Start Time [UTC]   & 18:08:38   & 00:52:25 && 09:30:359\\
Camera                   & ACIS-I-23/S123 & ACIS-I2-S123        && MOS1,2/pn               \\
Filter                   &  $--$                &       $--$                    && Medium  \\
Modes (read/data) & TIMED/VFAINT & TIMED/VFAINT && PFWE                 \\
Offset                   &   on-axis   & on-axis && on-axis     \\
Exposure [ks]        & 5.04   &         47.7    && 56.8-56.8/54.9         \\
GTI     [ks]                   & 4.97   & 47.1    && 52.1-52.3/43.9   &     \\
\hline
\end{tabular}
\end{center}
\tablefoot{All observations were taken from their respective mission archives. PFWE refers to the prime full window extended observation mode. The pointing of {\it Chandra} observations is centered on the following J2000.0 coordinates: $\alpha$= 13$^h$21$^m$50$\fs$89, $\delta$=-63$\degr$33$\arcmin$50$\farcs$0.}
\label{obstable}
\end{table*}

The canonical ways of  identifying the SN progenitors are i) to compare the ejecta abundances to the yields of SN models and ii) examine the morphology. Recently, \citet{lopez2011} have shown that the study of the X-ray morphology of SNRs can also be used to constrain their progenitors. In this case, the X-ray line and thermal  continuum emission of Type Ia SNRs tend to be more circular and mirror symmetric than core-collapse SNRs. The observed differences in morphologies are large enough to allow a clear separation of the two main explosion classes. This result is a consequence of the different geometries of the explosion mechanisms and circumstellar environments of Type Ia and core-collapse (CC) SNRs. Recently, \citet{peters2013} have extended the technique using ratios between different powers of a multipole expansion to analyze IR images of SNRs to test whether they can be classified using the symmetry of their warm dust emission as well.


\begin{figure*}
\centering
\includegraphics[width=6.3cm,angle=0]{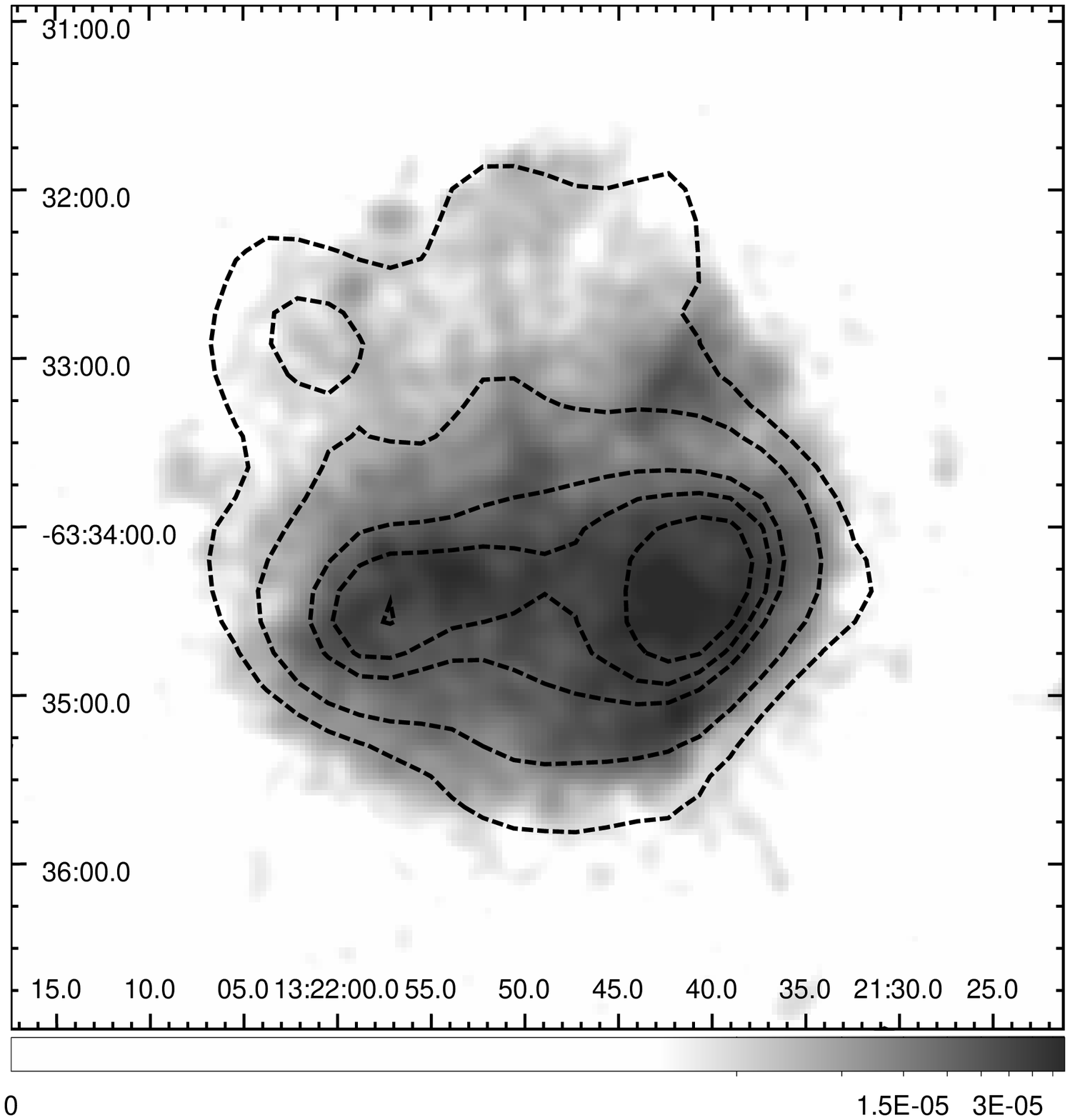}\hspace{-0.35cm}
\includegraphics[width=6.3cm,angle=0]{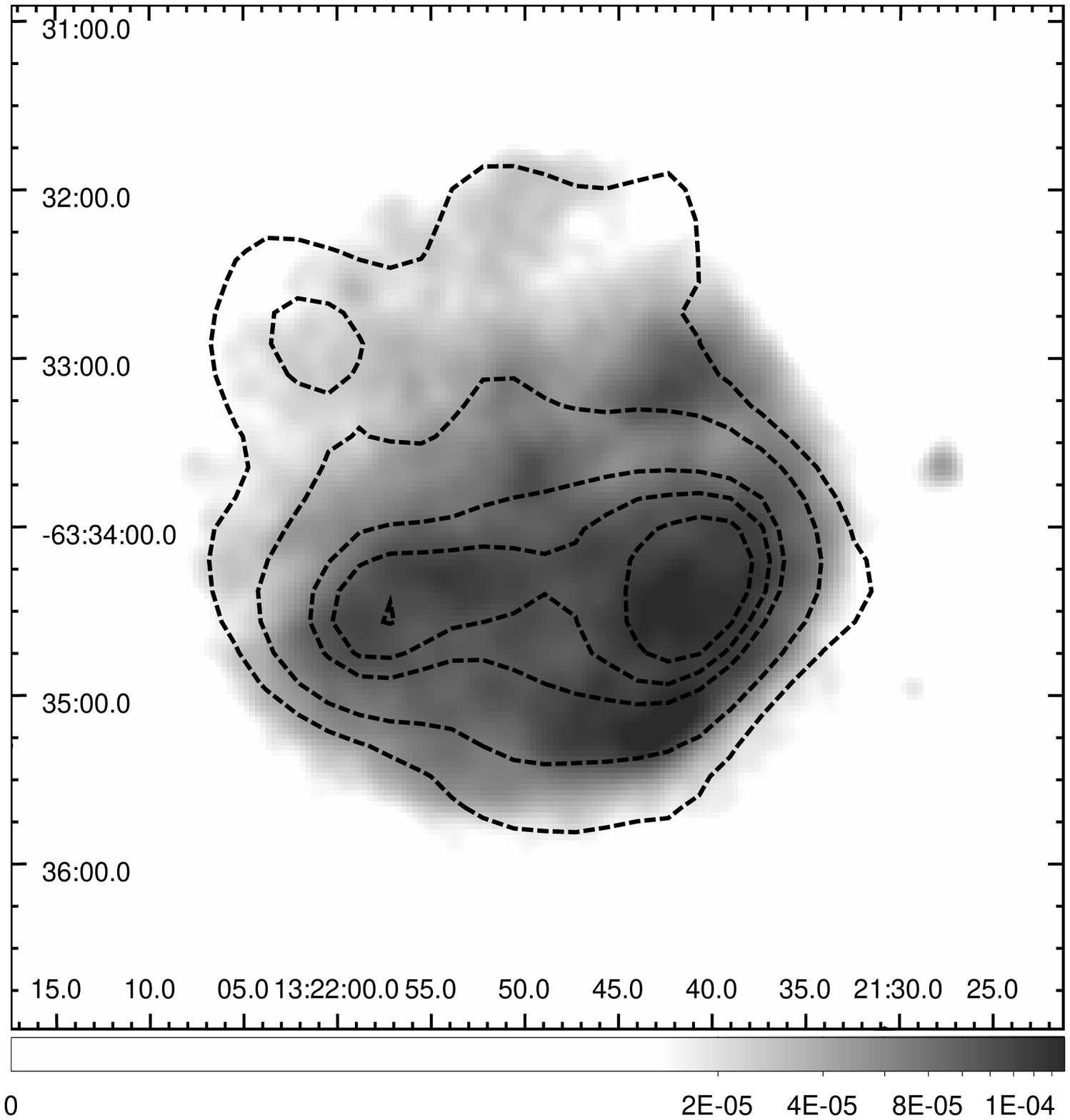}\hspace{-0.35cm}
\includegraphics[width=6.3cm,angle=0]{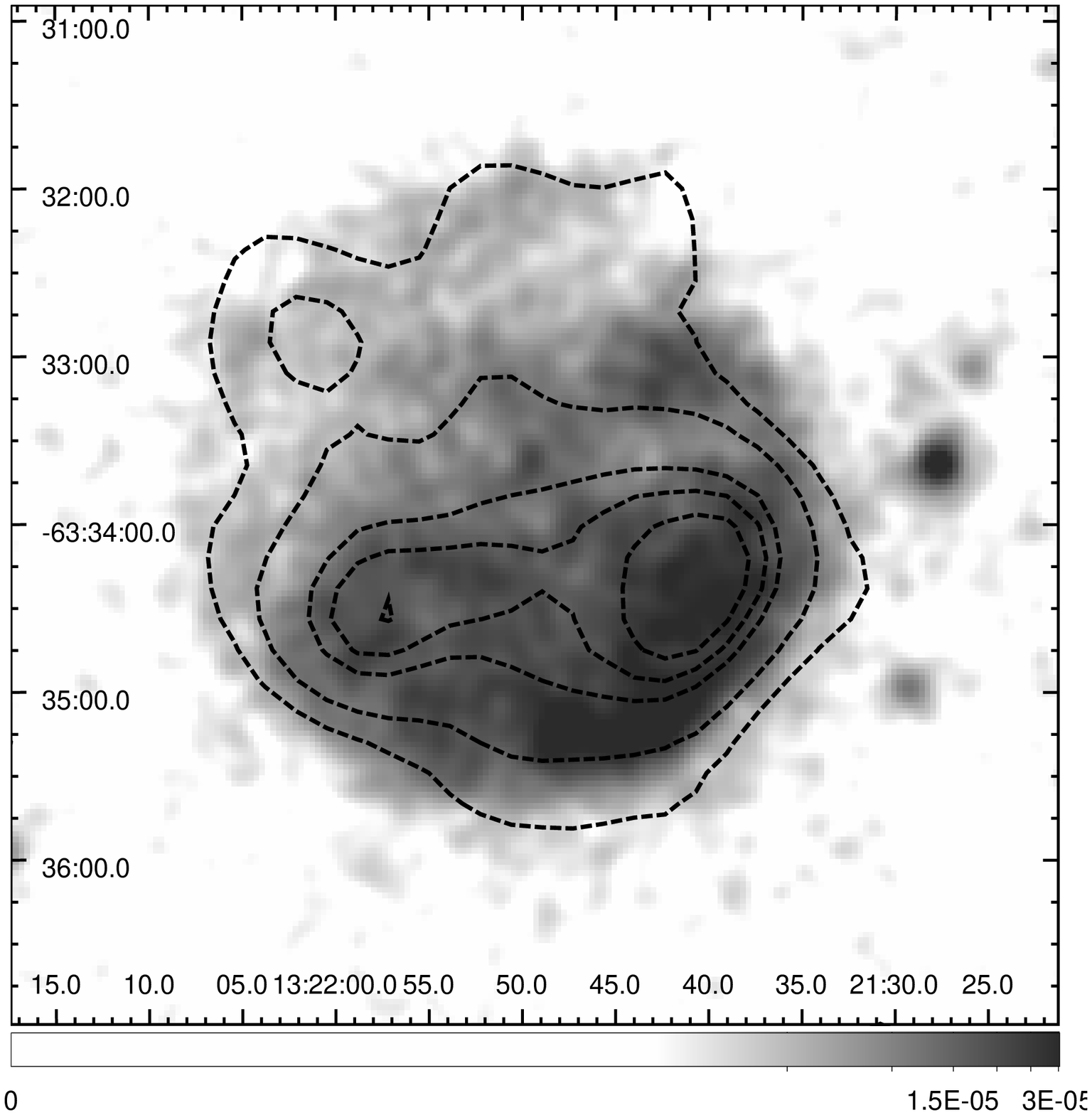}
\caption{XMM$-${\it Newton} X-ray images  with a size of 6.0 $\times$ 6.0 arcmin of SNR G306.3--0.9 in three energy bands. {\bf Left panel:} soft X-ray energies (0.5$-$1.0 keV). {\bf Middle:} medium X-ray energies (1.0$-$2.0 keV). {\bf Right:} hard X-ray energies (2.0$-$4.5 keV). Smoothed images were convolved with a 2D Gaussian function with a kernel of $\sigma =$ 20 arcsec. Overlapping contour levels at 2, 4, 10, 12, and 19 mJy~beam$^{-1}$ represent the 843~MHz radio image taken from the MOST Supernova Remnant Catalog \citep{whit96}.}
\label{tricolor}%
\end{figure*}

\begin{figure*}
\includegraphics[width=\textwidth,angle=0]{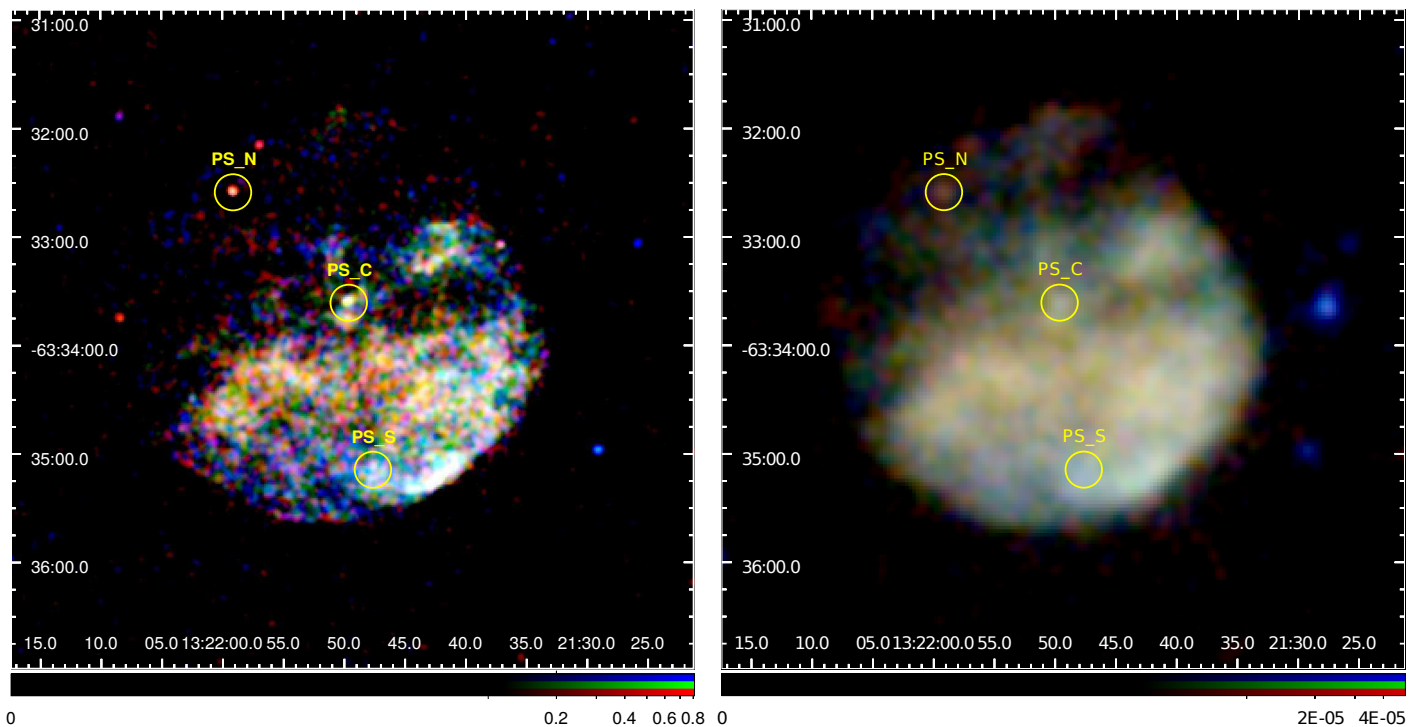}
\caption{{\bf Left panel:} {\it Chandra} image of G306.3$-$0.9 in three X-ray energy bands: soft (0.5$-$1.0 keV) in red, medium (1.0$-$2.0 keV) in green, and hard (2.0$-$4.5 keV) in blue. Soft X-ray point-like sources at the geometrical center of the SNR ($\alpha$= 13$^h$21$^m$50$\fs$2, $\delta$=$-$63$\degr$33$\arcmin$53$\farcs$9, J2000.0) and in the northeast part of the source are indicated as "PS C" and "PS N", respectively. A hard point-like source is indicated in the southern region as "PS S". {\bf Right panel:} Combined XMM-{\it Newton} PN, MOS1, and MOS2 image in the same energy ranges. Point-like sources are also indicated with individual yellow circles.}
\label{xmm-chandra}%
\end{figure*}

Therefore, with the information gathered from different observational facilities available such as XMM-{\it Newton}, {\it Chandra,} and {\it Spitzer,} it is possible to study a large number of SNRs in great detail \citep{vink2012} and to determine their progenitors. The southern Galactic SNR G306.3--0.9 that was recently
observed by the {\it Swift} telescope \citep{reynolds-atel}, is an interesting candidate to carry out this type of study, since the object displays a complex morphology at IR frequencies, dominated by a bright semicircular southern region \citep{reynolds2013}, which correlates very well with radio and X-ray emission observed by the {\it Chandra} telescope with its
high spatial resolution.

In this paper, we report on a detailed X-ray study of the SNR G306.3--0.9 using the enhanced sensitivity of the XMM-{\it Newton} and {\it Chandra} telescopes, which we complement with available
radio and IR data. The structure of the paper is as follows: in Sect. 2 we describe the XMM-{\it Newton} and {\it Chandra} observations and the data reduction. X-ray analysis and results are shown in Sect. 3. In Sect. 4 we discuss the implications of our results and, finally, we summarize our main conclusions in Sect. 5.

\section{X-ray observations and data reduction}

\begin{figure} 
\includegraphics[width=9.2cm,angle=0]{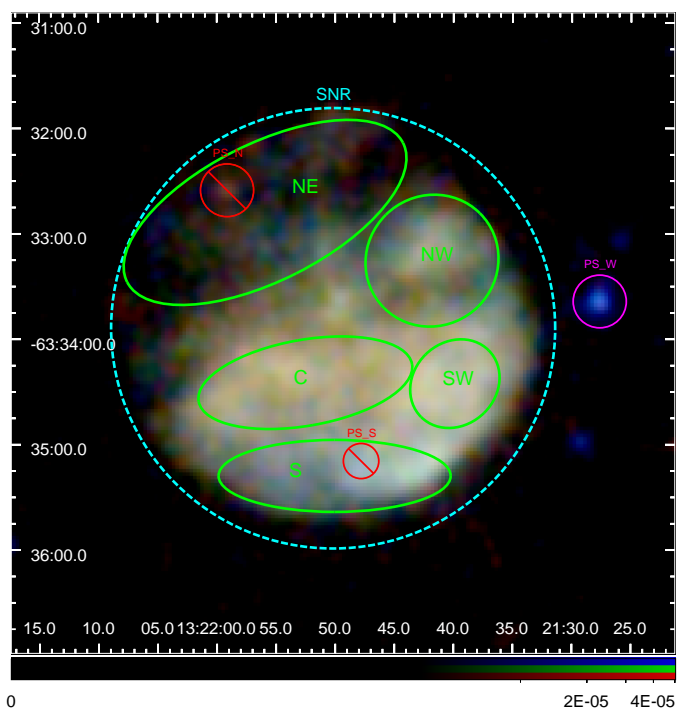}
\caption{XMM-{\it Newton} image, covering a 6$\times$6 arcmin$^{2}$ field of view, of SNR G306.3$-$0.9 in the three X-ray energy bands: soft (0.5--1.0 keV) in red, medium (1.0--2.0 keV) in green, and hard (2.0--4.5 keV) in blue. The boundary of the SNR is indicated in cyan. The selected X-ray spectra extraction regions are indicated in dark green for individual regions. Horizontal and vertical axes are labeled using J2000.0 right ascension and declination.}
\label{regiones}
\end{figure}

We combined XMM-{\it Newton} and {\it Chandra} data to carry out a detailed spectral and spatial X-ray analysis of SNR G306.3--0.9. The XMM-{\it Newton} observation was performed with the European Photon Imaging Camera (EPIC), which consists of three detectors, two MOS cameras \citep{turner2001}, and one PN camera \citep{struder2001}, operating in the 0.2$-$15~keV range. The satellite was pointed to $\alpha$= 13$^h$21$^m$44$\fs$45 and $\delta$=$-$63$\degr$34$\arcmin$34$\farcs$8 (J2000.0), with the SNR placed at the central CCDs. XMM-{\it Newton} data were analyzed with the Science Analysis System (SAS) version 14.0.0. Starting from level-1 event files, the latest calibrations (Master Index File = XMM$_{-}$CALINDEX$_{-}$0191.CCF) available at Sep-02-2015 in the CCF database, were applied with the {\sc emproc} and {\sc epproc} tasks. The events were then filtered to retain only photons likely for X-ray events: patterns 0 to 4 and energies 0.2 to 15.0 keV for the PN, patterns 0 to 12 and energies 0.2 to 12.0 keV for MOS1/2 instruments. To exclude strong background flares, we extracted light curves of the full field of view for each camera above 10~keV, excluding intervals 3-$\sigma$ above the mean count rate to produce good-time interval (GTI) files.

Two observations from the Advanced CCD Image Spectrometer (ACIS) camera are available in the {\it Chandra} archive. ACIS operates in the 0.1$-$10~keV range with high spatial resolution (0.5~arcsec). These observations were calibrated using the CIAO (version 4.7) and CALDB (version 4.6.7) packages by the {\sc chandra\_repro} task. Detailed information of the X-ray observations and the instrumental characteristics are given in Table~\ref{obstable}.

\begin{table*}
\caption{Spectral parameters of the diffuse X-ray emission of the selected regions.}
\renewcommand{\arraystretch}{1.3}
\begin{centering}
\begin{tabular}{l | c c c c c}
\hline\hline
Model \& Parameters & NE & NW & C & SW & S  \\
\hline
{\bf PHABS*(VAPEC+VNEI)} &&&&& \\
N$_\mathrm{H}$ [10$^{22}$~cm$^{-2}$]&    
1.57$^{+0.01}_{-0.02}$ &
1.54$^{+0.03}_{-0.01}$ &
1.53$^{+0.01}_{-0.01}$ &
1.55$^{+0.01}_{-0.01}$&
1.40$^{+0.03}_{-0.03}$ 
\\
kT$_{\rm VAPEC}$ [keV] &                                 
0.171$^{+0.006}_{-0.001}$   &
0.196$^{+0.002}_{-0.002}$   &
0.196$^{+0.002}_{-0.003}$   &
0.215$^{+0.001}_{-0.001}$   &
0.174$^{+0.008}_{-0.005}$ 
\\
Ne [Ne$_\odot$]        &
0.14$^{+0.06}_{-0.08}$ &
0.79$^{+0.12}_{-0.12}$ &
0.61$^{+0.05}_{-0.06}$&
0.66$^{+0.10}_{-0.09}$ &
0.55$^{+0.11}_{-0.11}$
\\
Mg [Mg$_\odot$]   &
0.23$^{+0.10}_{-0.09}$ &
0.50$^{+0.08}_{-0.08}$ &
0.32$^{+0.05}_{-0.05}$ &
0.35$^{+0.07}_{-0.07}$ &
0.49$^{+0.10}_{-0.20}$ 
 \\
Norm$_{\rm VAPEC}$ & 
0.12$^{+0.01}_{-0.02}$ &
0.039$^{+0.003}_{-0.010}$&
0.111$^{+0.002}_{-0.007}$ &
0.044$^{+0.006}_{-0.003}$ &
0.05$^{+0.01}_{-0.01}$
\\
\hline
kT$_{\rm VNEI}$ [keV] &                          
1.82$^{+0.15}_{-0.26}$   &
1.17$^{+0.07}_{-0.02}$   &
1.91$^{+0.05}_{-0.03}$   &
1.59$^{+0.10}_{-0.08}$   &
1.04$^{+0.05}_{-0.04}$  
\\
Si [Si$_\odot$] &    
8.68$^{+0.94}_{-2.16}$ &
1.79$^{+0.44}_{-0.15}$&
10.53$^{+1.01}_{-1.41}$  &
9.02$^{+11.24}_{-5.76}$ &
1.05$^{+0.09}_{-0.07}$  
\\
S [S$_\odot$] &  
12.50$^{+1.94}_{-3.05}$ &
3.11$^{+0.69}_{-0.31}$ &
18.42$^{+2.28}_{-2.30}$  &
16.47$^{+10.10}_{-3.33}$ &
1.43$^{+0.12}_{-0.10}$ 
\\
Ar [Ar$_\odot$] &  
11.92$^{+2.78}_{-3.84}$ &
2.10$^{+0.47}_{-0.63}$ &
13.41$^{+2.28}_{-2.30}$ &
18.69$^{+6.84}_{-9.07}$  &
1.51$^{+0.34}_{-0.53}$  
\\
Ca [Ca$_\odot$] &  
19.81$^{+8.50}_{-7.41}$&
4.51$^{+2.01}_{-1.54}$ &
28.62$^{+6.47}_{-6.45}$ &
23.66$^{+6.84}_{-9.08}$ &
2.58$^{+0.88}_{-0.94}$ 
\\
Fe [Fe$_\odot$]    & 
18.56$^{+2.85}_{-4.63}$ &
4.09$^{+1.09}_{-0.50}$ &
46.25$^{+1.98}_{-2.91}$  &
34.29$^{+16.24}_{-12.47}$  &
1.05$^{+0.14}_{-0.12}$  
 \\
$\tau$[$10^{10}$~s~cm$^{-3}$] &
3.61$^{+0.68}_{-0.28}$&
9.25$^{+2.08}_{-1.53}$ &
3.76$^{+0.23}_{-0.12}$ &
5.89$^{+0.51}_{-0.35}$ &
9.67$^{+1.87}_{-1.15}$ 
\\
Norm$_{\rm VNEI}$ [$10^{-3}$] & 
0.09$^{+0.01}_{-0.01}$ &
0.48$^{+0.08}_{-0.04}$&
0.093$^{+0.003}_{-0.007}$ &
0.075$^{+0.028}_{-0.024}$ &
1.47$^{+0.10}_{-0.14}$
\\
\hline
$\chi^{2}_{\nu}$ / d.o.f. &
1.16 / 558 &
1.13 / 568 &
1.36 / 681 &
1.27 / 548 &
1.12 / 634 
 \\
\hline
Flux(0.5$-$1.0~keV)& 
0.49$^{+0.05}_{-0.06}$&
0.60$^{+0.03}_{-0.09}$&
1.38$^{+0.23}_{-0.24}$&
0.69$^{+0.01}_{-0.21}$&
0.63$^{+0.02}_{-0.04}$
\\
Flux(1.0$-$2.0~keV) & 
2.62$^{+0.42}_{-0.32}$&
3.45$^{+0.05}_{-0.67}$&
6.05$^{+1.19}_{-0.87}$&
3.57$^{+0.02}_{-1.36}$&
4.02$^{+0.01}_{-0.17}$
\\
Flux(2.0$-$7.5~keV) & 
1.53$^{+0.58}_{-0.05}$&
1.72$^{+0.08}_{-0.37}$&
2.85$^{+1.14}_{-0.96}$&
1.79$^{+0.02}_{-0.77}$&
2.43$^{+0.02}_{-0.09}$
\\
\hline
Total Flux(0.5$-$7.5~keV)&
4.65$^{+0.89}_{-0.66}$&
5.77$^{+0.25}_{-0.77}$&
10.30$^{+2.95}_{-2.60}$&
6.05$^{+0.04}_{-2.36}$&
7.09$^{+0.01}_{-0.26}$ \\
\hline
\end{tabular}
\label{allspectable}
\tablefoot{Normalization is defined as 10$^{-14}$/4$\pi$D$^2\times \int n_H\,n_e dV$, where $D$ is distance in [cm], n$_\mathrm{H}$ is the hydrogen density [cm$^{-3}$], $n_e$ is the electron density [cm$^{-3}$], and $V$ is the volume [cm$^{3}$]. $\chi^{2}_{\nu}$ / d.o.f is the reduced chi-squared. Error values are 1$\sigma$ (68.27\%) confidence intervals for each free parameter and unabsorbed fluxes are given in units of 10$^{-13}$~erg~cm$^{-2}$~s$^{-1}$. Abundances are given relative to the solar values of \cite{anders1989}.}
\end{centering}
\end{table*}

\section{Results}
\subsection {X-ray images}

To perform the morphological analysis and select the regions for the spatially resolved spectroscopy, we combined X-ray images of EPIC MOS and PN to increase the signal-to-noise ratio (S/N) by means of the {\sc emosaic} SAS task to merge the images. The corresponding set of exposure maps for each camera was prepared to account for spatial quantum efficiency and mirror vignetting by running the SAS task {\sc eexmap}. Exposure vignetting corrections were performed by dividing the merged count image by the corresponding merged exposure maps. 

Because of the high spatial resolution and sensitivity of the data set, we were able to examine the X-ray morphology of the supernova remnant in detail. In Fig.~\ref{tricolor} we show narrow-band images generated in the energy ranges 0.5$-$1.0~keV, 1.0$-$2.0~keV, and 2.0$-$4.5~keV using the EPIC MOS 1/2 cameras, with superimposed  843~MHz radio contours \citep{whit96}. In the image, North is up and East is to the left. The SNR shows a semi-circular and asymmetric X-ray morphology, with diffuse X-ray emission predominantly located in the southern part of the remnant, and very weak extended X-ray emission in the northern region. In the southern region the images also reveal a typical rim-brightened outer SNR shell, as first noted by \citet{reynolds2013}.

The left-hand panel of Fig.~\ref{xmm-chandra} shows a composite three-color image of the {\it Chandra} observations in three energy bands: soft (0.5$-$1.0 keV) in red, medium (1.0$-$2.0 keV) in green, and hard (2.0$-$4.5 keV) in blue. The remnant is undetected above 4.5~keV in {\it Chandra} data. The right-hand panel of Fig.~\ref{xmm-chandra} shows the XMM-{\it Newton} X-ray image in the same energy band for comparison. In this XMM-{\it Newton} observation the remnant is detectable up to $\sim$8.0~keV. Smoothed images were obtained by the convolving with a 2D Gaussian with a kernel of $\sigma =$ 20~arcsec. The XMM-{\it Newton} and {\it Chandra} images reveal similar characteristics. The sensitivity of the {\it Chandra} observations allowed us to detect, for the first time, three possible X-ray point-like sources. One located near the geometrical center of the X-ray structure of the SNR at $\alpha$= 13$^h$21$^m$50$\fs$2, $\delta$=$-$63$\degr$33$\arcmin$53$\farcs$9, J2000.0 (indicated in Fig.~\ref{xmm-chandra} as PS~C), another embedded within weak diffuse emission near the northeast edge of the remnant (indicated as PS~N), and the third located in the southern part of the SNR (indicated as PS~S), which is embedded in bright X-ray emission.
\begin{figure} 
\includegraphics[width=0.45\textwidth]{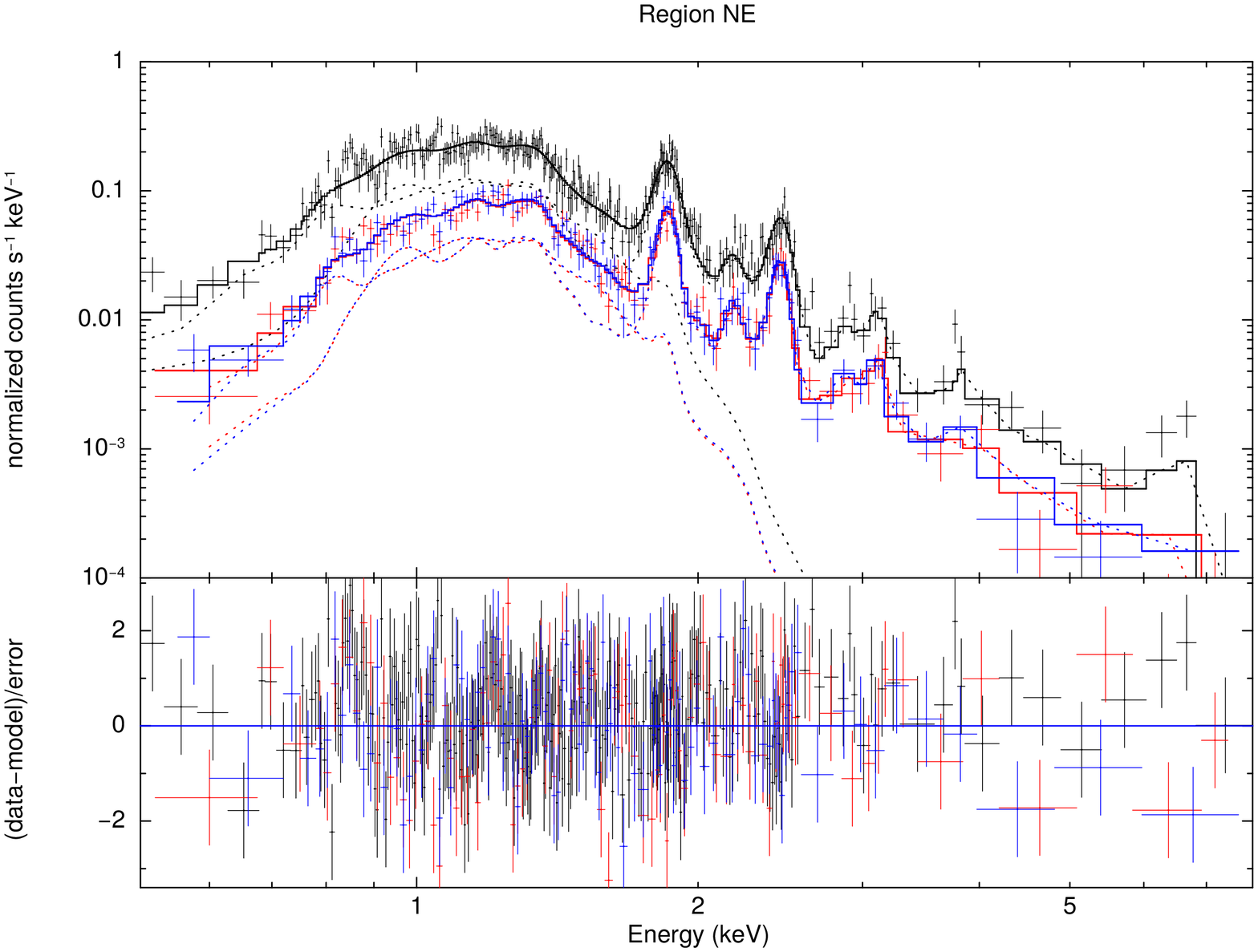} ~~~~
\includegraphics[width=0.45\textwidth]{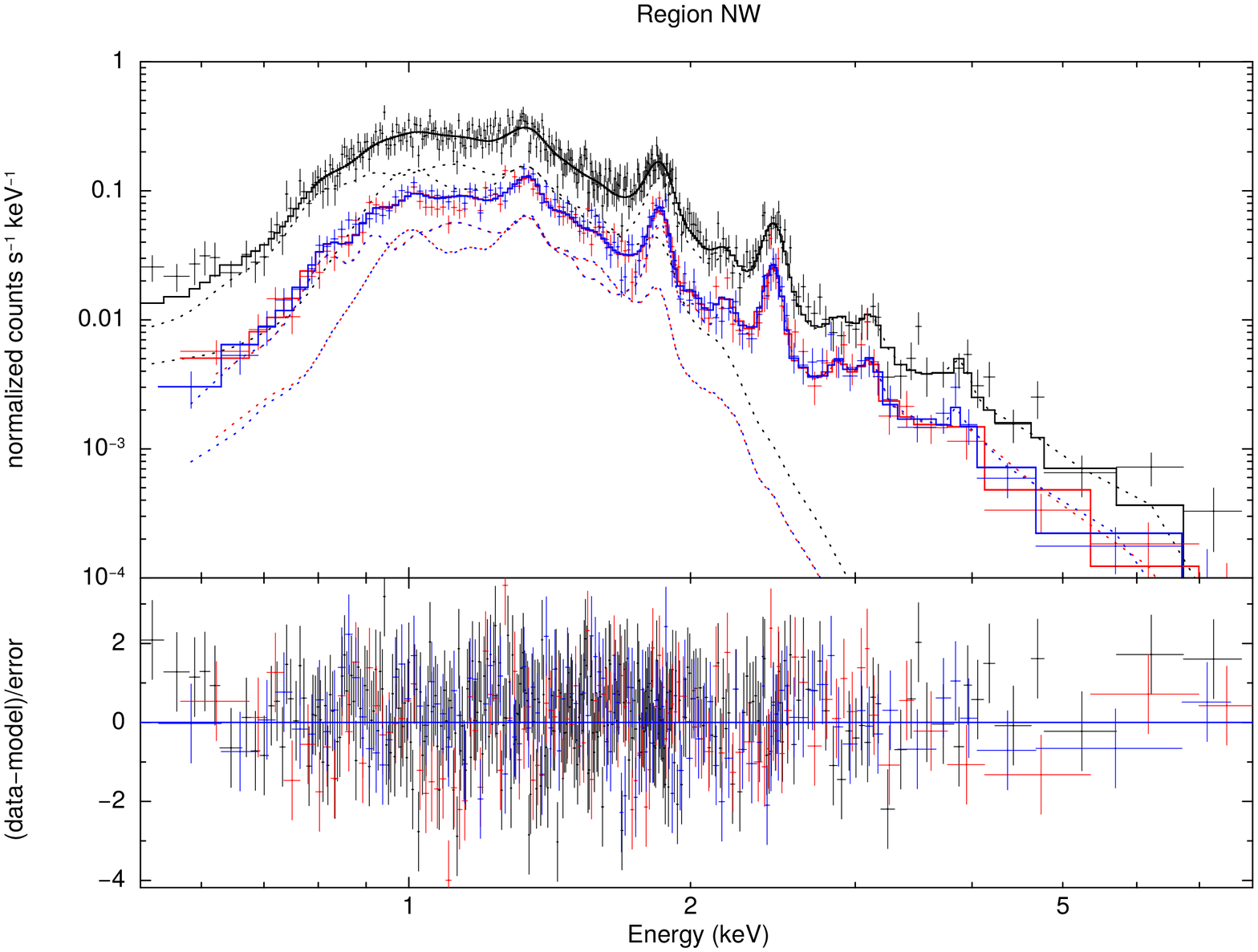} ~~~~
\includegraphics[width=0.45\textwidth]{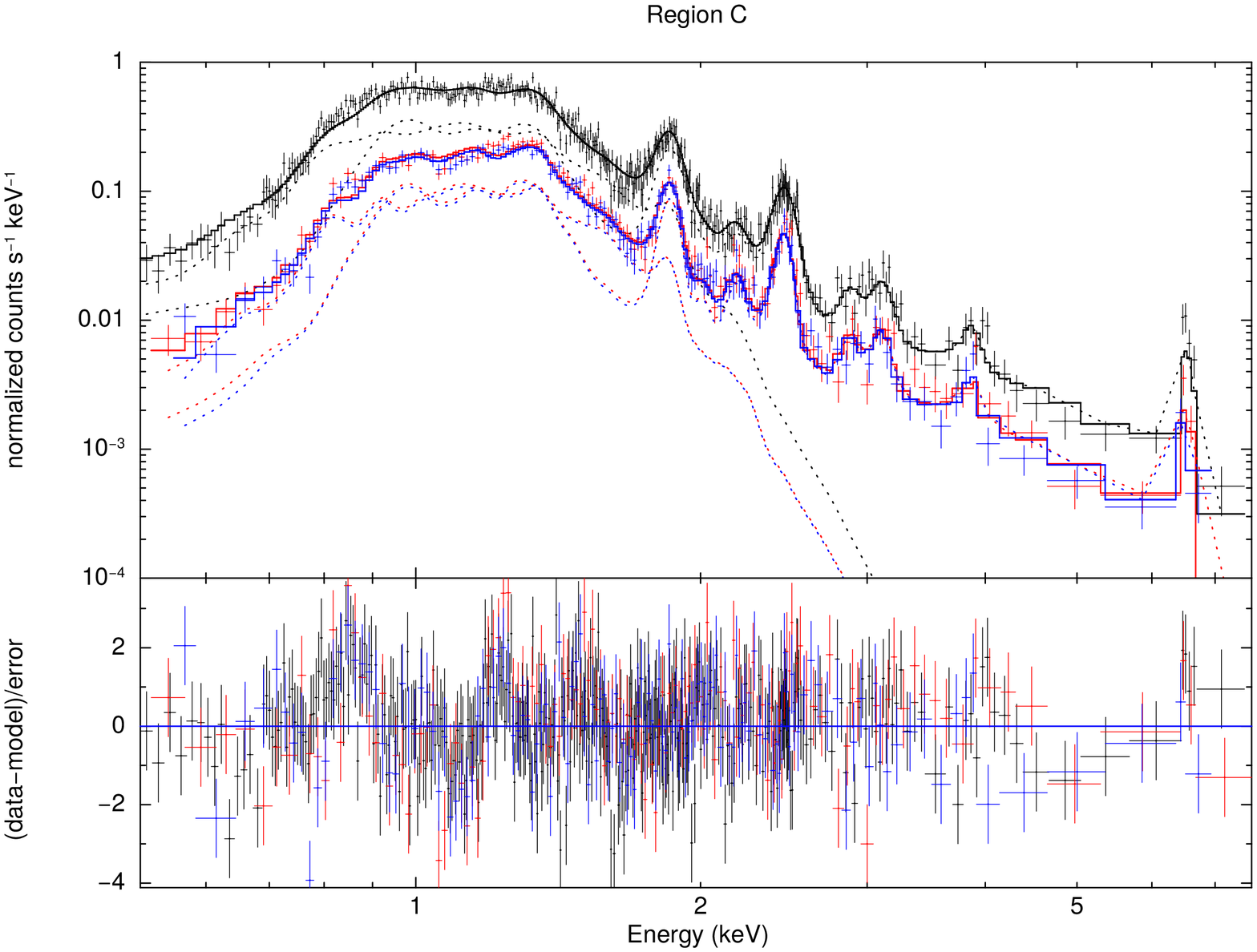}
\caption{XMM-{\it Newton} PN and MOS1/2 spectra of NE, NW, and C regions. Solid lines indicate the best-fit two-component model (see Table~\ref{allspectable}). Dashed lines indicate individual contributions of each VAPEC and VNEI thermal plasma models for each camera. Lower panels present the $\chi^2$ fit residuals.}
\label{r1-r4e}%
\end{figure}

\begin{figure} 
\includegraphics[width=0.45\textwidth]{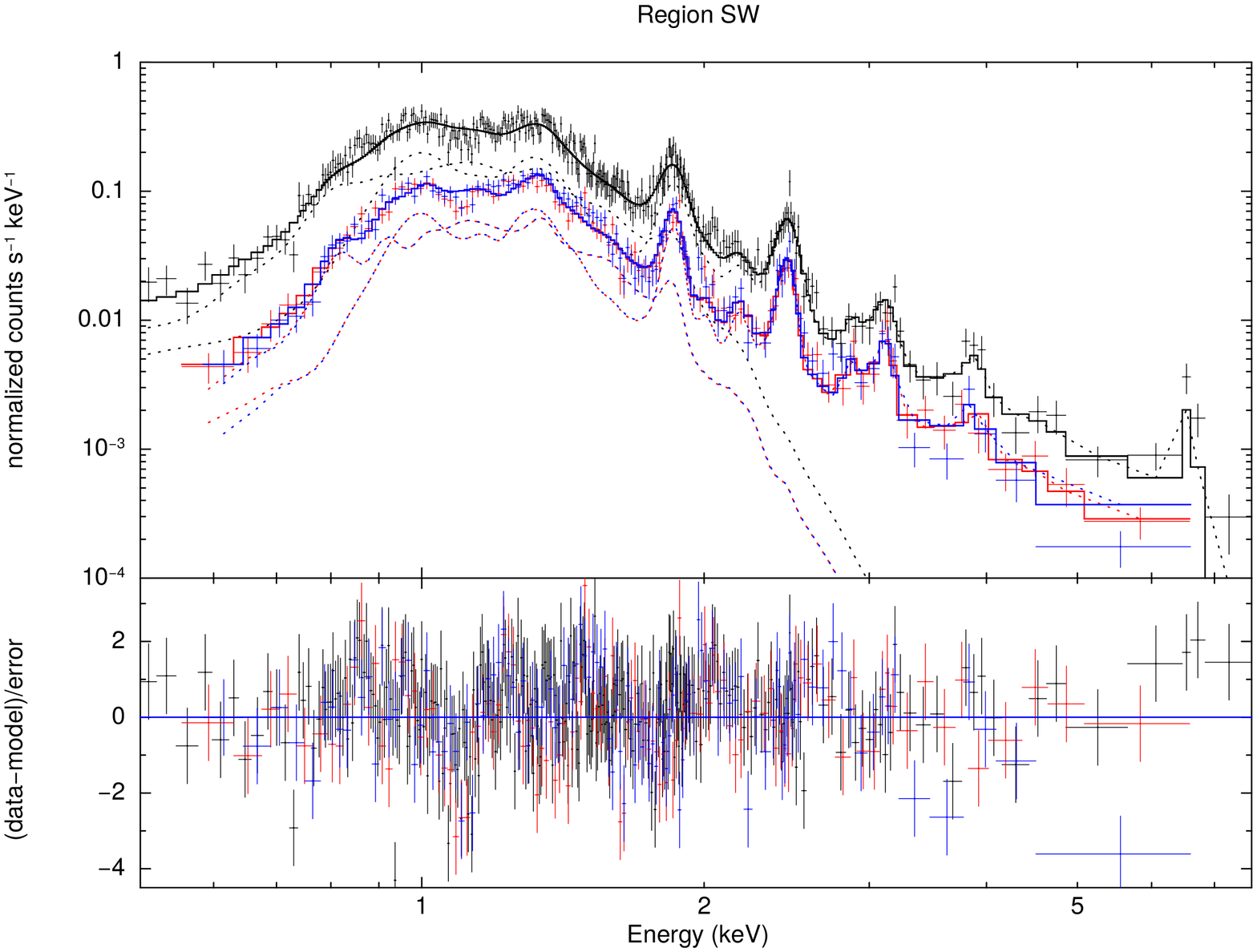}
\includegraphics[width=0.45\textwidth]{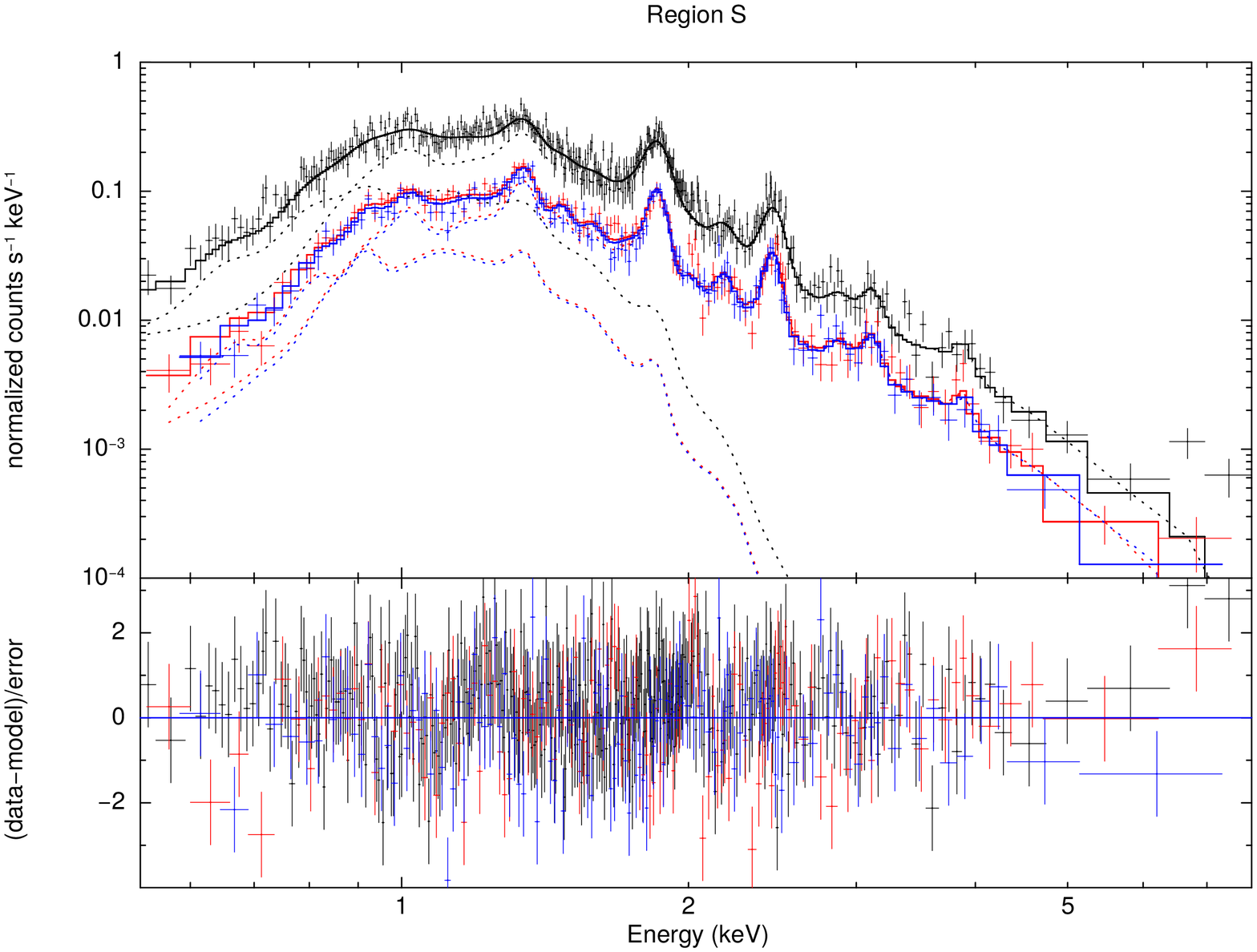}
\caption{XMM-{\it Newton} PN and MOS1/2 spectra of the SW and S regions. Solid lines indicate the best-fit two-component model (see Table~\ref{allspectable}). Dashed lines indicate individual contributions of each VAPEC and VNEI thermal plasma models for each camera. Lower panels present the $\chi^2$ fit residuals.}
\label{r4w-r6}%
\end{figure}

The central X-ray source is located at $\alpha_{\rm J2000.0}$=$13^{\rm h} 21^{\rm m} 49\fs9 $, $\delta_{\rm J2000.0}$=$-63\degr 33\arcmin 37\farcs2 $ (hereafter we refer to this object as CXOU J132149.9$-$633337). From the {\it Chandra} data, a total of 83 counts were detected in a single event file in the 0.5$-$4.5 keV using a radius of 2 arcsec. The X-ray source has a $\sim$ 7 $\sigma$ significance above local background, which is a quite robust detection. The source PS N is located at $\alpha_{\rm J2000.0}$=$13^{\rm h} 21^{\rm m} 59\fs2 $, $\delta_{\rm J2000.0}$=$-63\degr 32\arcmin 35\farcs1$, and has a $\sim$ 25 $\sigma$ significance above local background. Finally, the source PS~S at $\alpha_{\rm J2000.0}$=$13^{\rm h} 21^{\rm m} 47\fs8 $, $\delta_{\rm J2000.0}$=$-63\degr 35\arcmin 07\farcs8$, presents a $\sim$ 11 $\sigma$ significance above local background. 

From the analysis of the X-ray images we found that the source PS~N is dominant in the soft X-ray band and is coincident with the IR source 2MASS J13215916$-$6332255 \citep{cutri2003}; the source PS~S is dominant in the hard X-ray band and displays X-ray characteristics typical of an active galactic nucleus; and the source PS~C is dominant in the medium X-ray band and is marginally coincident with the IR source 2MASS J13214960-6333360 \citep{cutri2003}. Unfortunately, the low-resolution radio data do not allow observing any point-like counterpart. All of them are absent in the 843~MHz radio map.

To check whether CXOU J132149.8$-$633335 is a point-like object, we searched for extended X-ray emission (i.e., a pulsar wind nebula) that might be associated with the source. For this purpose, we applied the {\sc srcextent} tool of the CIAO software package. As a result, the analysis shows that when we use radii $\ga$5~arcsec, the source extends to sizes of about 5~arcsec. Conversely, when we use a radius of 4~arcsec, its size is 1.6~arcsec, which agrees with the PSF computed at the position of the source.  

\subsection{X-ray spectral analysis}

To analyze physical and chemical conditions of the X-ray emission in SNR G306.3--0.9 in detail, we extracted spectra from regions with different sizes, chosen on the basis of the morphology observed in the X-ray images (see Fig.~\ref{regiones}). After several tests, we found that the best choice included five individual regions (NE, NW, C, SW, and S), which are indicated in green in Fig.~\ref{regiones}. Spectra were obtained using the {\sc evselect} SAS task with the appropriate parameters for EPIC MOS~1/2 and PN cameras. Background spectra were extracted from circular regions with radii of 1.75" adjacent to the SNR, but where it does not emit X-rays. For all regions, bright point-like sources were excluded, as indicated in Fig.\ref{regiones}. Consistent results were obtained through analyzing Chandra data, but only the XMM-{\it Newton} spectra are shown for simplicity.

Figures \ref{r1-r4e} and \ref{r4w-r6} show the XMM-{\it Newton} background-subtracted X-ray spectra obtained for the different regions of the SNR. In these figures, the spectra are grouped with a minimum of 16~counts per bin. Error values are 1$\sigma$ (68.27\%) confidence levels for each free parameter and $\chi^{2}$ statistics are used. The spectral analysis was performed using the XSPEC package (Version 12.9.0) \citep{arnaud1996} and the  emission line information from AtomDB database (Version 3.0.2). 

The spectra of regions were fit using different models: APEC, NEI, VNEI, PSHOCK, and VPSHOCK, each modified by an absorption interstellar model \citep[PHABS;][]{balucinska1992}. After several tests, we found that the best fit for the individual regions is consistent with a VAPEC with sub-solar abundances of Ne and Mg, and a VNEI model \citep{borkowski2001}, dominated by elevated abundances of Si, S, Ar, Ca and Fe. It is interesting to note that the central region C and the SW region show a strong Fe line at $\sim6.4$~keV, typical of ejecta material, which is also present in the NE region. This two-component plasma model has two electron temperatures, one associated with ejecta material with a hot temperature kT$_{\rm VNEI}$ and another related to swept-up ISM medium with a lower temperature kT$_{\rm VAPEC}$. The X-ray parameters of the best fit to the diffuse emission spectra for the different regions are presented in Table~\ref{allspectable}.

\begin{figure} 
\centering
\includegraphics[width=9.0cm,angle=0]{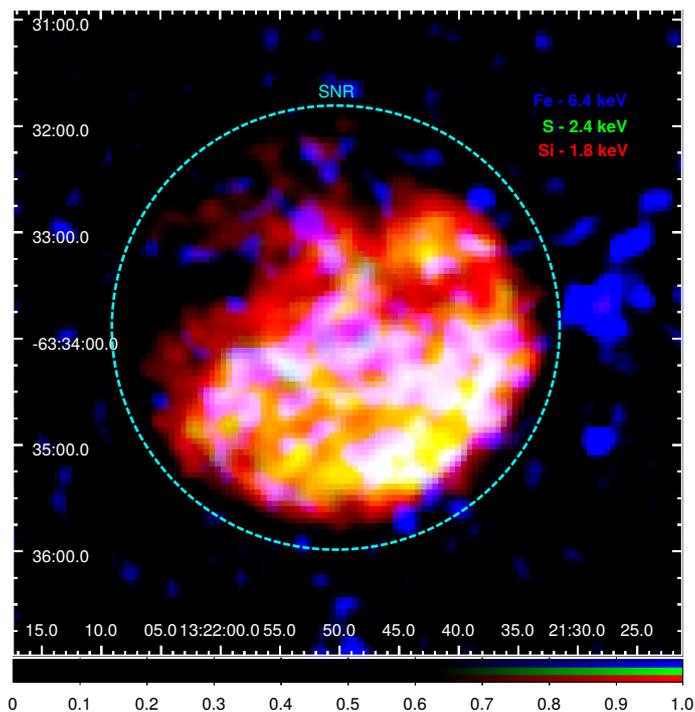}
\caption{Equivalent width maps obtained from the merged EPIC event files of the XMM-{\it Newton} for Si XIII (1.85 keV, in red), S XV (2.4 keV, in green) and Fe-K (6.4 keV, in blue). Cyan contour indicates the peripheral edge of total X-ray emission of G306.3--0.9. Horizontal and vertical axes are labeled using J2000.0 right ascension and declination.}
\label{fig-lines}
\end{figure}

\begin{figure} 
\centering
\includegraphics[width=8.7cm,angle=0]{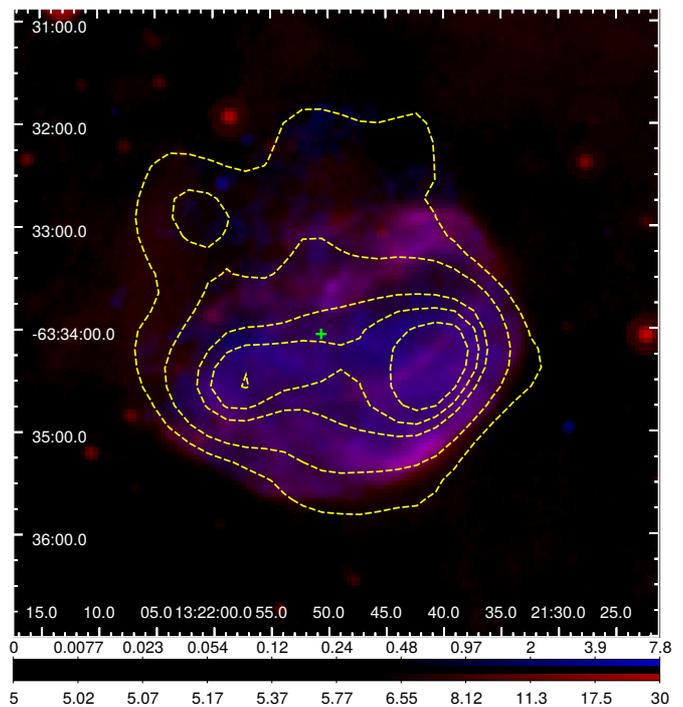}
\caption{Composite image of the radio (dotted yellow contours), IR (red color) from MIPS-{\it Spitzer} in mJy~sr$^{-1}$ and X-ray (blue color) emission of SNR G306.3$-$0.9. The green cross indicates the IR centroid of the SNR, used in the PRM analysis (see text for details). Horizontal and vertical axes are labeled using J2000.0 right ascension and declination.} 
\label{fig-spitzer}%
\end{figure}

The spatially resolved spectral analysis shows that the physical conditions of the plasma are not homogeneous throughout the remnant. In spite of this, the values of column density ($N_{\rm H}$) are consistent between each other within the errors, ranging from 1.41 $\times$ 10$^{22}$ cm$^{-2}$ to 1.57$\times$ 10$^{22}$ cm$^{-2}$. Lower values were obtained for the regions NW, SW, and S, while the highest values of $N_{\rm H}$ correspond to regions NE and C. Moreover, the values obtained for the temperature $kT_{\rm VAPEC}\sim0.19$ suggest that the ISM is at similar conditions throughout the SNR. 

In the ejecta model, two different sets of regions are distinguished. Regions NE, C, and SW show significantly higher temperatures and larger abundances than NW and S regions, which seems to indicate that the ejecta in the central and northern part of the SNR are rich in heavy elements while in regions closer to the main shock front the metals are less abundant, thus suggesting a higher concentration of heavy elements in the remnant interior. The relatively short ionization timescale ($\lesssim 10^{11}$~cm$^{-3}$~s) in all the regions confirms that the ejecta component is far from the equilibrium ionization throughout the source. Regions NW and S display relatively higher values of the ionization timescale than regions NE, C, and SW, whose ionization timescales are shorter, all of them being $\sim 4 \times 10^{10}$~cm$^{-3}$~s. The VNEI component displays temperatures $\sim$1.1~keV and long ionization timescales for the NW and S external regions, with moderate abundances of Si, S, Ar, Ca, and Fe, and temperatures in the 1.6--1.9~keV range and shorter ionization timescales for NE, C and SW regions, with enhanced abundances of those elements, indicating that the ejecta is reheated as a consequence of a reverse shock acting on the inner regions of the SNR. 

The spatial variation of Si, S, and Fe in the global morphology of the SNR is also observed in the equivalent width maps shown in Fig.~\ref{fig-lines}. This image was obtained using narrow energy ranges of 1.75--1.95 keV, 2.3--2.6 keV, and 6.2--6.8 keV for Si (in red), S (in green), and Fe (in blue) lines, respectively. The maps reflect the non-uniformity in the spatial distribution of the elements, which are mostly concentrated in the southern semicircular region. To obtain the equivalent widths for each of the XMM-{\it Newton} EPIC cameras, we followed a standard procedure. We obtained images in wider continuum bands of (1.5--2.1 keV) for Si, (2.1--2.8 keV) for S and (5.5--7.5) for Fe, which include the narrow line bands described above. We scaled them by a factor corresponding to the ratio of the flux in that band to the flux in the line band from the continuum spectral model assuming the results obtained for the central region. Using this number, we obtained images corresponding to the continuum below each line, which we then subtracted from the line images to keep only the line contribution. The equivalent width maps show more clearly the stratification of Fe concentrated in a ridge of emission toward the center of the remnant, as has been noted by \cite{reynolds2013}, which is also coincident with the bright ridge observed at radio wavelengths by MOST. None of the point-like sources detected within the SNR display typical characteristics of the so-called central compact objects \citep[CCO, see][for a review on their properties]{halpern2010} according to our spectral X-ray analysis.

Finally, it is interesting to note that although our fit for the individual regions is consistent with those from \cite{reynolds2013} using {\it Chandra} observations, the spatially resolving spectral analysis allowed us to pinpoint inhomogeneities in the X-ray emitting plasma. Therefore, it is evident that large-exposure observations and the simultaneous use of the improved effective area from XMM-{\it Newton} and spatial resolution from {\it Chandra} plays an important role in mapping small-scale spectral changes in this SNR.

\section {Discussion}

The X-ray morphology and global spectral properties of SNR G306.3$-$0.9 were first investigated in previous studies using limited {\it Chandra} observations \citep{reynolds2013}. Thanks to the large spectral effective area of the instruments onboard XMM-{\it Newton}, together with a deeper {\it Chandra} observation with high spatial resolution and {\it Spitzer} data available at the archive, we were able to develop a detailed morphological and spatially resolved spectral X-ray analysis, which is required to outline a realistic astrophysical scenario that can describe the remnant's evolution and to understand the properties derived from a multiwavelength study.

Our results show that the X-ray morphology of the remnant displays several interesting characteristics, which comprise a southern semicircular brightened structure of X-ray emission with a northern part of fainter emission. Its semicircular global morphology is also consistent with radio and IR observations, as previously noted by \citet{reynolds2013}. 

In Fig.~\ref{fig-spitzer} we present the mid-IR emission at 24 $\mu$m of G306.3$-$0.9. A comparison with the X-ray images shows a strong correlation of the two emissions, as is usually found in SNRs \citep{morton07,arendt10,williams11,dubner13}. The 24 $\mu$m emission is dominated by thermal continuum emission of dust, most likely very small, stochastically heated dust grains \citep{draine03}. The abundance of small grains in SNRs is probably due to the dust shattering generated by the shock fronts \citep{andersen11}, which then are heated by electrons and X-ray photons within the hot thermal gas.

On the other hand, the fact that in the central part of the remnant the Fe abundance is high favors the possibility of a Type Ia SN origin. This is also consistent with the S/Si, Ar/Si, and Ca/Si abundances ratio (1.75, 1.27, and 2.72 in the central region, respectively), which are broadly in accordance with a degenerate scenario and are highly inconsistent with any core-collapse model \citep{nomoto1997}. Moreover, following \cite{yamaguchi2014}, we obtained the centroid of the Fe-K line in the central region of the SNR. As a result, by fitting a power law to the continuum emission in the 5--8~keV energy range plus a Gaussian for the emission line, we estimated a centroid of $6.52\pm0.01$~keV, which also agrees with a Type Ia scenario. This possibility can be independently supported using the morphological analysis proposed by \citet{lopez2011} and \citet{peters2013}, who used the asymmetry of X-ray and 24~$\mu$m IR emissions to classify the SNR as Type
Ia or core-collapse (CC). This technique is independent of the plasma conditions and easy to perform. It is based on the power-ratio method (PRM), which gives quantitative values of asymmetries in images by obtaining the multipole moments of the corresponding surface brightness in a circular aperture \citep[see][for a detailed description]{lopez2011}. 

When we assume that the origin of the moments lies in the centroid of the image (see green cross in Fig.~\ref{fig-spitzer}, which corresponds to Spitzer data), the dipole power approaches zero, but the normalized quadrupole power $P_2/P_0$ measures the ellipticity of the distribution, while the octupole power $P_3/P_0$ characterizes the mirror asymmetry of the source. Following \cite{lopez2011}, in our case we used the soft band (0.5-2.1 keV) and \ion{Si}{xiii} emission line (1.75--2.0 keV) for X-ray images of SNR G306.3--0.9 from {\it Chandra} ACIS camera. We also applied the method to the 24~$\mu$m IR image from the Multiband Imaging Photometer (MIPS; \citet{riecke2004}) in the {\it Spitzer} telescope. Background was extracted and the multipole powers were calculated referring to the centroid of each image (differences in X-ray centroids are within the 0.5 arcsec ACIS resolution, while the distance between centroids in the IR and X-ray bands is $\sim$15~arcsec). As a result, $P_2/P_0 = 2.48 \times10^{-6}$ and $P_3/P_0 = 8.03\times 10^{-8}$ were calculated for the X-ray soft band, while $P_2/P_0 = 2.00 \times10^{-6}$ and $P_3/P_0 = 7.97\times 10^{-8}$ were obtained for the \ion{Si}{xiii} line. For the 24 $\mu$m image, we obtained $P_2/P_0 = 8.54 \times10^{-6}$ and $P_3/P_0 = 4.38\times 10^{-7}$. These values are consistent with a Type Ia SNR (see Fig. 4 in \citet{peters2013} and Fig. 2 in \citet{lopez2011}). 

Finally, using the information obtained from the spectral analysis, we can roughly compute the electron density of the plasma and age of the different regions of the SNR. Assuming that the plasma fills ellipsoidal regions like that  used to extract the spectra (indicated in Fig.~\ref{regiones}), we obtained a volume $\bar{V}$ for each region of the SNR for distances in the 6$-$10~kpc
range. Thus, based on the emission measure (EM) determined by the spectral fitting (see Table 2), we estimated the plasma electron density by $n_{\rm e}$=$\sqrt{EM/V}$. As a result, assuming a filling factor, $f \equiv 1$, and considering an average between the elliptical semi-axes for the unknown length of the ellipsoids along the line of sight, for the regions NE, NW, C, SW, and S we obtained mean electron densities of 0.31, 1.12, 0.60, 0.85, and 2.75~cm$^{-3}$, respectively. In addition, considering a sphere with the same radius as the projected SNR and using the length of the chord intercepted by each ellipsoidal region, we obtain upper limits for their volumes and thus lower limits for their densities of 0.19, 0.69, 0.43, 0.42, and 1.41~cm$^{-3}$ for the NE, NW, C, SW, and S regions, respectively. In all cases, the number density of the nucleons was simply assumed to be the same as that of the electrons. Hence, despite all the uncertainties inherent to these calculations, the densities estimated in the NW and S external parts of the SNR are higher than those found for the internal and N regions. Using the ionization timescale $\tau$ of Table 2, we found that the elapsed time $t=\tau /n_{\rm e}$ after the plasma was heated is $\sim$ 2300 $\pm$ 1300 yr. These values agree very well with the age obtained by Reynolds et al. (2013), adopting a Sedov model (Sedov 1959).

\section{Conclusions}

We have used deep XMM-{\it Newton} and {\it Chandra} observations of SNR G306.3--0.9 to study the properties of its X-ray emission and its physical connection with radio and IR observations in
detail. The global X-ray morphology reveals a semi-circular X-ray shape, with clear stratifications of Fe concentrated in a ridge of emission toward the center of the remnant. This emission is coincident with the bright filament observed at IR wavelengths. 

Spectral analysis confirmed that the physical conditions of the emitting plasma are inhomogeneous throughout the remnant. The X-ray spectral analysis of the individual regions showed emission-line features of metal elements, confirming the thermal origin of the emission. The X-ray spectra are reasonably well represented by two absorbed VAPEC and VNEI thermal plasma models, with temperatures in the range of 0.17--0.21 keV for the VAPEC model,  corresponding to emission arising from the shocked ISM and a VNEI component with temperatures $\sim$1.1~keV and long ionization timescales for the NW and S external regions, with moderate abundances of Si, S, Ar, Ca, and Fe, and temperatures in the 1.6--1.9~keV range and shorter ionization timescales for NE, C, and SW regions, with enhanced abundances of those elements, indicating that the ejecta has been excited to X-ray emitting temperatures by a reverse shock acting on the inner regions of the SNR. The values of abundances found in the central region of G306.3--0.9 and the centroid of the Fe-K line favor a Type Ia progenitor for this remnant. The outer regions show values of the abundances above solar but lower than to those found in the central regions. This suggests that the composition is a combination of ejecta and shocked ISM material. Moreover, an independent morphological study performed on the IR and {\it Chandra} X-ray data also suggests a Type Ia origin.

Deeper dedicated observations both in radio and X-ray wavelengths are required to search for a compact remnant immersed in the bright thermal plasma emission from this young SNR.

\begin{acknowledgements}
We thank the anonymous referee for the insightful comments and constructive suggestions that lead to an improved manuscript. JAC and SP are CONICET researchers. J.A.C was supported on different aspects of this work by  Consejer\'{\i}a de Econom\'{\i}a, Innovaci\'on, Ciencia y Empleo of Junta de Andaluc\'{\i}a under excellence grant FQM-1343 and research group FQM-322, as well as FEDER funds. S.P. is supported by CONICET, ANPCyT and UBA (UBACyT) grants. FG and AES are fellows of CONICET.
\end{acknowledgements}


\end{document}